\documentstyle[preprint,aps,amsfonts]{revtex}
\tighten
\begin{document}
\preprint{Preprint Numbers: 
\parbox[t]{50mm}{ADP-93-215/T133\\
		 ANL-PHY-7599-TH-93\\
		 FSI-SCRI-93-108}}
\draft
\title{Dynamical chiral symmetry breaking and confinement \\
with an infrared-vanishing gluon propagator?}

\author{Frederick T. Hawes\footnotemark[2], 
Craig D. Roberts\footnotemark[1] and 
Anthony G. Williams\footnotemark[3]\footnotemark[4]\vspace*{2mm} }

\address{\footnotemark[2]Department of Physics and SCRI, 
Florida State University, \\ 
Tallahassee, Florida 32306-3016\vspace*{2mm}\\
\footnotemark[1]Physics Division, Argonne National Laboratory,
Argonne, Illinois 60439-4843 \vspace*{2mm}\\ 
\footnotemark[3]Department of Physics and Mathematical Physics, \\
  University of Adelaide, \\
  Adelaide 5001, South Australia, Australia}

%
\maketitle
%
\begin{abstract}
We study a model Dyson-Schwinger equation for the quark propagator closed
using an {\it Ansatz} for the gluon propagator of the form 
\mbox{$D(q) \sim q^2/[(q^2)^2 + b^4]$} and two {\it Ans\"{a}tze} for 
the quark-gluon vertex: the minimal Ball-Chiu and the modified form suggested
by Curtis and Pennington.  Using the quark condensate as an order parameter,
we find that there is a critical value of $b=b_c$ such that the model does
not support dynamical chiral symmetry breaking for $b>b_c$.  We discuss and
apply a confinement test which suggests that, for all values of $b$, the
quark propagator in the model {\bf is not} confining.  Together these results
suggest that this Ansatz for the gluon propagator is inadequate as a model
since it does not yield the expected behaviour of QCD.
\end{abstract}
%
\section{Introduction}
Dynamical Chiral Symmetry Breaking (DCSB) and confinement are two crucial
features of quantum chromodynamics (QCD).  Indeed, it might be argued that a
realistic model of QCD should manifest both of these properties since they
are responsible for the nature of the hadronic spectrum; DCSB ensuring the
absence of low mass scalar partners of the pion and confinement ensuring the
absence of free quarks, for example.

A natural method for studying both DCSB and confinement in QCD, and models
thereof, is the complex of Dyson-Schwinger Equations (DSEs).~\cite{DSERef}
The equations for the two-point functions of gluons and quarks have been used
in many such studies.  This manifestly relativistically covariant approach,
recent reviews of which can be found in Refs.~\cite{H91,CDR93}, has provided
the foundation for a useful and successful understanding of the phenomena of
low energy QCD by facilitating the construction of realistic field theoretic
models.~\cite{CDM}

Another goal in the DSE studies is to develop this nonperturbative approach
to a point where it is as firmly founded as lattice QCD and calculationally
competitive.  Although much remains to be done in order to achieve this goal
there has been a good deal of progress, especially in the study of Abelian
gauge theories where direct and meaningful comparisons can be made, and
agreement obtained, between the results of lattice and DSE
studies.\cite{QEDCJB}

In considering DSE studies it is important to note that they are hampered by
the fact that there is an infinite tower of coupled equations: the equations
for the two-point functions couple them to other two- and three-point
functions; the equations for the three-point functions couple them to the
four-point and higher n-point functions; etc.  A commonly used resolution of
this problem is to truncate the system at a finite number of coupled
equations by making {\it Ans\"{a}tze} for the higher n-point functions.  For
example, one may study the DSE for the quark propagator alone by choosing
{\it Ans\"{a}tze} for the gluon propagator and quark-gluon vertex thus
closing the system.  This is the nature of our study.

Herein we study the fermion DSE obtained with a model gluon propagator
(2-point function) which vanishes at $q^2=0$:
\begin{equation}
\label{SProp}
      D(q) \equiv \frac{1}{q^2[1-\Pi(q^2)]} = \frac{q^2}{(q^2)^2 + b^4}~,
\end{equation}
where $\Pi(q^2)$ is the gluon vacuum polarisation function and $b$ is a real
parameter, in order to determine whether it can support DCSB and/or generate
a confining quark propagator.  Two properties we require of the propagator of
a confined particle are: 1) the absence of a K\"{a}llen-Lehmann
representation and 2) no singularity on the timelike $q^2$ axis.\cite{RWK92}
The gluon propagator obtained from Eq.~(\ref{SProp}) satisfies both of these
requirements.

Such a form for the gluon propagator, even though it may be argued to
describe a confined gluon, is perhaps counterintuitive, since it would appear
to provide a weak interaction between quarks at small $p^2$, which
corresponds to large distances.  Indeed, some studies of the fermion DSE in
QCD have employed quite a different {\it Ansatz}: one which behaves as
$1/(q^2)^2$ for \mbox{$q^2\simeq 0$}.~\cite{BP88,SAA91} This form for the
infrared (IR) behaviour of the gluon propagator is suggested by a number of
studies of the DSE for the gluon propagator in both axial~\cite{GIRA} and
covariant gauges~\cite{GIRC,UG80} using various approximation and/or
truncation procedures; notably, they all effectively neglect the 4-gluon
vertex.  (We note that the results of the axial gauge studies~\cite{GIRA} may
be questioned on the basis that the gluon propagator therein is inconsistent
with the known spectral representation in axial gauge.~\cite{GW83}) In
addition, this form of gluon propagator in the infrared is consistent with
area law behaviour of the Wilson loop~\cite{GW82} which has been observed in
lattice gauge theory studies of QCD~\cite{latQCD} and is often regarded as
indicating confinement.

However, the form in Eq.~(\ref{SProp}) is suggested by a number of studies.
It has been argued~\cite{Zw91} that in order to completely eliminate Gribov
copies,~\cite{VG79} and hence to fix Landau or Coulomb gauge uniquely in
lattice studies, one must introduce new ghost fields into QCD in addition to
those associated with the Fadde'ev-Popov determinant in the continuum.
Analysing the lattice action thus obtained suggests that the gluon propagator
vanishes as $(q^2)^\gamma$, with $\gamma>0$ not determined.  Subsequent
analysis of a simplified model yields $\gamma=1$ and, in fact, a gluon
propagator of the form in Eq.~(\ref{SProp}) with $b$ a finite constant in
Landau gauge.  Similar considerations in Ref.~\cite{VG79} yield the same
result.  A propagator of the form in Eq.~(\ref{SProp}) was also suggested in
Refs.~\cite{St86,St90a,St90b} as a result of an analysis of an approximate
DSE for the gluon propagator. A recent lattice QCD simulation~\cite{CBern}
also provides some support for this form.  We also note that a gluon
propagator of the type in Eq.~(\ref{SProp}) may arise in the field strength
approach to QCD.~\cite{FSA}

It is therefore important to study the phenomenological implications of
Eq.~(\ref{SProp}); i.e, to determine whether it can support DCSB and
confinement in QCD based models of the type in Refs.~\cite{CDM}.  It has been
argued~\cite{Zw91,St86,St90a,St90b} that Eq.~(\ref{SProp}) represents
confined gluons because there are no poles on the timelike real axis in the
complex-$q^2$ plane and it allows the interpretation of the gluon as an
unstable excitation which fragments into hadrons before observation (in a
time of the order of $1/b$).  It is also argued~\cite{St90b} that such a
gluon propagator should lead to a quark propagator with similar structure in
the complex plane, and hence a similar interpretation, but this result has
not been proven.  Learning sufficient about the quark propagator to make
inferences about its analytic structure is therefore an important part of our
study.

We study the implications of Eq.~(\ref{SProp}) for the structure of the quark
propagator and DCSB using the fermion DSE.  A similar study is undertaken in
Ref.~\cite{Axel93}.  Our DSE is closed using two {\it Ans\"{a}tze} for the
quark-gluon vertex: the minimal Ball-Chiu vertex~\cite{BC80} and that of
Curtis and Pennington.~\cite{CPcoll} These vertices are free of kinematic
light-cone singularities.  We find that in either case there are regions of
DCSB and unbroken chiral symmetry characterised by a two-dimensional phase
diagram in $(b^2,\ln\tau)$ space, where $b$ appears in Eq.~(\ref{SProp})
above and $\tau$ will be introduced below in connection with an ultraviolet
modification of Eq.~(\ref{SProp}). The phase transition is second order.  We
also employ a confinement test which suggests that, with either of the vertex
{\it Ans\"{a}tze} and for all values of $b$ and $\tau$, Eq.~(\ref{SProp})
leads to a quark propagator that {\bf is not} confining.  Given these results
it appears that the gluon propagator of Eq.~(\ref{SProp}) is inadequate as as
a model, at least in our DSE framework, since it does not yield the behaviour
expected in QCD.

In Sec.~II we present a model quark DSE which we solve numerically.  We also
discuss the gluon propagator and quark-gluon vertex {\it Ans\"{a}tze} in some
detail.  In Sec.~III we evaluate the quark condensate, which we use as an
order parameter for DCSB, and determine the characteristic properties of the
phase transition.  We also demonstrate that the quark propagator is not
confining.  In Sec.~IV we discuss and summarise our results and conclusions.

\section{Model Dyson-Schwinger Equation}

In Minkowski space, with metric \mbox{$g_{\mu\nu} = \;{\rm diag}
(1,-1,-1,-1)$} and in a general covariant gauge, the inverse of the dressed
quark propagator can be written as
\begin{equation}
 S^{-1}(p) = \not\!{p} - m - \Sigma(p)
	   \equiv Z^{-1}(p^2) \left( \not\!{p} -M(p^2) \right)
           \equiv A(p^2) \not\!{p} - B(p^2)\/,
\end{equation}
with $m$\ the renormalised explicit chiral symmetry breaking mass (if
present), $\Sigma(p)$\ the self-energy, $M(p^2) = B(p^2)/A(p^2)$\ the
dynamical quark mass function and $A(p^2) = Z^{-1}(p^2)$ the
momentum-dependent renormalisation of the quark wavefunction.

The unrenormalised DSE for the inverse propagator is
\begin{equation}
 S^{-1}(p) = \not\!{p} - m^{\rm bare}
     - i \case{4}{3} g^2 \int \frac{d^4k}{(2\pi)^4} \gamma^\mu
         S(k) \Gamma^\nu (k,p) D_{\mu \nu}((p-k)^2)\/,     \label{fullDSE}
\end{equation}
where $\Gamma^\nu$ is the proper quark-gluon vertex and $D_{\mu\nu}(q^2)$\ is
the dressed gluon propagator.  Hereafter we set
\mbox{$m^{\rm bare} = 0$} so that we can concentrate on dynamical
symmetry breaking effects.  (A study of explicit chiral symmetry breaking in
such DSE models of QCD can be found in Ref.~\cite{WKR91,HW91}.)  The
renormalised, massless DSE is
\begin{equation}
 S^{-1}_R(p) = Z_S\not\!{p} 
     - i Z_\Gamma \case{4}{3}  g^2 \int^{\Lambda}
         \frac{d^4k}{(2\pi)^4} \gamma^\mu
        S_R(k) \Gamma^\nu_R (k,p) D_{\mu \nu}^R((p-k)^2)\/,     \label{fullRDSE}
\end{equation}
where $Z_S$ and $Z_\Gamma$ are quark-propagator and quark-gluon-vertex
renormalisation constants, respectively, which depend on the renormalisation
scale, $\mu$, and ultraviolet cutoff, $\Lambda$.  Neglecting ghost fields and
explicit 3-gluon vertices, as we do in the following (see
Eq.~(\ref{Piapprox}) below), one has \mbox{$Z_S=1=Z_\Gamma$} at one-loop in
Landau gauge.  Using this result here leads to our model renormalised DSE for
the quark self energy:
\begin{equation}
 \Sigma(p)=   i \case{4}{3}  g^2 \int^{\Lambda}
         \frac{d^4p}{(2\pi)^4} \gamma^\mu
         S(k) \Gamma^\nu (k,p) D_{\mu \nu}((p-k)^2)\/,     \label{ModRDSE}
\end{equation}
where here and hereafter we suppress the label $R$.  

\subsection{Model Gluon Propagator}
In a general covariant gauge the dressed gluon propagator, which is diagonal
in colour space, can be written:
\begin{equation}
\label{Gprop}
D^{\mu\nu}(q^2) = \left[ \left(g^{\mu\nu}-\frac{q^\mu q^\nu}{q^2}\right)
  \frac{1}{1-\Pi(q^2)} - a \frac{q^\mu q^\nu}{q^2} \right] 
\frac{1}{q^2}\/,
\end{equation}
where $\Pi(q^2)$\ is the gluon vacuum polarisation and $a$\ is the gauge
parameter.  It should be noted that in covariant gauges the longitudinal
piece of this propagator is not modified by interactions.  This follows from
the Slavnov-Taylor indentities in QCD.  (See, for example, Ref.~\cite{PT84},
pp.  42-45.)

The form of the propagator we consider herein has been argued to arise in
Landau gauge ($a=0$) and we shall use that gauge hereafter.  (This has the
benefit that the gluon propagator is purely transverse.  In addition, Landau
gauge is a fixed point of the renormalisation group and therefore the gauge
parameter does not ``run''.  [See, for example, Ref.~\cite{PT84}, pp.
135-136.])

A commonly used approximation is to write 
\begin{equation}
\label{Piapprox} 
\frac{g^2}{4\pi} \frac{1}{1-\Pi(q^2)} = \alpha(-q^2)
\end{equation}
where $\alpha(-q^2)$\ is the running coupling in the gauge theory.  In QCD
this amounts to neglecting ghost contributions to the gluon vacuum
polarisation.~\cite{UG80} (In Abelian theories, of course, this is not an
approximation but an exact result which follows from the fact that, in the
absence of ghost fields and gauge-boson self-interactions, the
renormalisation constants for $\Pi(q^2)$ and $g^2$ are the same.\cite{IZ80})  
For $Q^2 = -q^2 \gg \Lambda_{QCD}^2$\/, the one loop contribution is dominant
and the running coupling is
\begin{equation}
  \alpha_{\rm S}(Q^2) = \frac{d\pi}{\ln(Q^2/\Lambda_{QCD}^2)}~;
\;\;\;\; d = \frac{12}{33-2N_f}.
\end{equation}
(We use $N_f=4$ herein.  Our results are insensitive to changes in this
value.) In QCD, using Eq.~(\ref{Piapprox}) in Eq.~(\ref{Gprop}) is expected
to be accurate at large $Q^2$ where it leads to the correct perturbative
leading-log behaviour.~\cite{QCDLL}

In considering the small $q^2$ behaviour we return to Eq.~(\ref{SProp}).  We
observe that in Refs.~\cite{St86,St90a,St90b} a solution of the coupled DSEs
for the gluon propagator and triple-gluon vertex was obtained, using rational
polynomial {\it Ans\"{a}tze}, (the coupling to the 4-gluon vertex was
eliminated) and it was argued that Eq.~(\ref{SProp}) is a good approximation
to the solution.  A significant result in this study is that, in order to
support a solution of the type in Eq.~(\ref{SProp}), the transverse part of
the triple-gluon vertex, \mbox{$\Gamma_{3T}(p_1,p_2,p_3)$}, necessarily has
kinematic light-cone singularities of the form $1/(p_i^2)$.  These
singularities, which cannot arise in perturbation theory, are introduced in
``solving'' the Slavnov-Taylor identities for the triple-gluon vertex.  In
this connection we remark that Ref.~\cite{BC80} might be argued to suggest
that the Slavnov-Taylor identities for both the triple-gluon vertex and
quark-gluon vertex can be solved without introducing kinematic light-cone
singularities.  This latter approach would allow for nonperturbative
corrections to the vertices and a simple connection with the perturbative
vertices in the asymptotically free region.  In this case the light-cone
singularities in the triple-gluon vertex of Refs.~\cite{St90a,St90b} might be
viewed as undesirable.

We note too that there has been an attempt to employ the gluon propagator
described by Eq.~(\ref{SProp}) in a study of quarkonium
spectra.~\cite{BLPSR91} A Blankenbecler-Sugar reduction of a ladder-like
approximation to the Bethe-Salpeter equation is used and it is argued that in
the bound state equation that results one can approximate the effect of
Eq.~(\ref{SProp}) by a Coulomb potential for all $r$.  The interesting
structure in this equation arises because the bound fermions are described by
propagators with complex conjugate poles just as the gluon propagator has
(see Eq.~(\ref{FStingl}) and associated discussion below).  The results in
this study, however, are not competitive with detailed fits using potential
models.  It may be that the approximations/truncations made in
Ref.~\cite{BLPSR91} are partly responsible for this.

In connection with the small $q^2$ behaviour of the quark-quark interaction
the lattice results of Ref.~\cite{CBern} are also interesting.  This study,
on $16^3\times 40$ and $24^3\times 40$ lattices at $\beta=6.0$, yielded a
gluon propagator in a lattice Landau gauge which allowed a fit of the form in
Eq.~(\ref{SProp}) at small $q^2$ but which could not rule out a fit using a
standard massive particle propagator.  Other lattice sizes and values of
$\beta$ were also studied.  The results at $\beta=6.3$ on a lattice of
dimension $24^4$ were not inconsistent with these results but in this case
the small physical size of the lattice was a problem.  On a lattice of
dimension $16^3\times 24$ at $\beta=5.7$ it was found that the gluon
propagator was best fit with a standard massive vector boson propagator with
mass $\sim 600$~MeV.  (We note that the gauge fixing in this study did not
include the modifications suggested by Ref.~\cite{Zw91}.)  These studies
represent an improvement in both technique and lattice sizes over earlier
lattice studies of the gluon propagator~\cite{MO87} but the conclusions are
not markedly different.  The studies of Ref.~\cite{MO87}, using $\beta= 5.6,
6.0$ on a $4^3\times 8$ lattice and $\beta= 5.8$ on a $4^3\times 10$ lattice,
obtained results that were consistent with a free massive boson propagator
with mass $\sim 600$~MeV.  Clearly, further lattice studies would be of great
interest.

At present, one can only say that
Refs.~\cite{Zw91,VG79,St86,St90a,St90b,CBern} suggest that in QCD
\begin{equation}
\frac{1}{1-\Pi(q^2)} = \frac{(q^2)^2}{(q^2)^2 + b^4}
\label{resb}
\end{equation}
is not implausible, at least at small $q^2$.  For this reason we study
Eq.~(\ref{SProp}) in the DSE approach in order to determine whether it can
lead to DCSB and a confining quark propagator.  This will provide further
insight into the validity of this form of gluon propagator.

Combining Eqs.~(\ref{Piapprox}) and (\ref{resb}) leads to the
ultraviolet-improved model gluon propagator that we consider herein; i.e, we
study the phenomenological implications of a model gluon propagator obtained
with
\begin{equation}
\label{PiA}
\frac{g^2}{4\pi} \;\frac{1}{1-\Pi(q^2)} = \alpha(\tau; -q^2)\;
\frac{(q^2)^2}{(q^2)^2 + b^4}~, 
\end{equation}
where (\mbox{$Q^2 \equiv -q^2$})
\begin{equation}
\label{alphatau}
\alpha(\tau; Q^2) = 
\frac{d\pi}
{\ln\left[\displaystyle\tau + \frac{Q^2}{\Lambda_{\rm QCD}^2}\right]}~,
\end{equation}
in Eq.~(\ref{Gprop}).  Here \mbox{$\tau > 1$} is an IR regularisation
parameter introduced so that the logarithmic singularity is shifted to
\mbox{$Q^2= -\tau \Lambda_{\rm QCD}^2$} which ensures that the piece derived
from Eq.~(\ref{resb}) dominates in the spacelike IR region.~\cite{AJ88,RM90}

\subsection{Model Quark-gluon Vertex}
In choosing an {\it Ansatz} for the vertex we note that in QCD the
Slavnov-Taylor identity~\cite{MP78}
\begin{equation}
\label{STI}
q_\mu\Gamma^\mu(k,p)\left[1+b(q^2)\right] =
  \left[1-B(q,p)\right]S^{-1}(k) - S^{-1}(p)\left[1- B(q,p)\right]~,
\end{equation}
where $q=(p-k)$, $b(q^2)$ is the ghost self energy and $B(p,q)$ is the
ghost-quark scattering kernel, constrains the longitudinal part of the
vertex.  Clearly, the often used ``rainbow'' approximation, in which
$\Gamma_\mu(k,p)$ is simply taken to be $\gamma_\mu$, cannot satisfy
Eq.~(\ref{STI}) in an interacting theory.

Neglecting ghosts, as we have done so far, this relation takes the form of
the Ward-Takahashi identity in QED:
\begin{equation}
\label{WI}
q_\mu\Gamma^\mu(k,p)= S^{-1}(k) - S^{-1}(p)~.
\end{equation}
The constraints that this relation place on the vertex can now be inferred
from the QED studies.~\cite{BC80,CPcoll,BR93} Taking these into account one
is lead to a vertex of the form
\begin{eqnarray}
\label{VA}
\Gamma^{\mu}(k,p)
  & = & \Gamma^\mu_{\rm BC}(k,p) + \Gamma^\mu_{\rm CP}(k,p)\/,
\end{eqnarray}
where
\begin{eqnarray}
\Gamma^{\mu}_{\rm BC}(k,p)
  & = & \frac{A(p^2)+A(k^2)}{2} \gamma^\mu        \label{VBC} \\
  &   & + \frac{(p+k)^{\mu}}{p^2 -k^2}
        \left\{ \left[ A(p^2)-A(k^2)\right]
                  \frac{\left[ \not\!{p} + \not\!{k} \right]}{2}
        - \left[ B(p^2) - B(k^2) \right] \right\}\/,    \nonumber \\
\Gamma^\mu_{\rm CP}(k,p) & = &
\frac{\gamma^\nu(k^2-p^2) - (k+p)^\nu (\not\!{k}-\not\!{p})}{2d(k,p)}
      \left[ A(k^2)-A(p^2) \right]\/,             \label{CPGamT} \\
  {\rm with}~d(k,p) & =& \frac{1}{(k^2+p^2)}\left( (k^2-p^2)^2
	   + \left[ \frac{B^2(k^2)}{A^2(k^2)}+\frac{B^2(p^2)}{A^2(p^2)}
	      \right]^2 \right)~.    \label{CPdkp}
\end{eqnarray}
In these equations, $\Gamma_{\rm BC}$ is the Ball-Chiu vertex of
Ref.~\cite{BC80} and $\Gamma_{\rm CP}$ is the additional piece suggested and
studied by Curtis and Pennington in Refs.~\cite{CPcoll}.

Equation~(\ref{VA}) specifies our vertex {\it Ansatz} which has the properties
that it satisfies the Ward-Takahashi Identity, is free of kinematic
singularities, reduces to the bare vertex in the absence of interactions,
transforms correctly under charge conjugation and Lorentz transformations and
preserves multiplicative renormalisability in the quark DSE.  (Of these
properties the minimal Ball-Chiu vertex satisfies all but the last.) 

For the most part in the following we neglected $\Gamma^\mu_{\rm CP}(k,p)$;
i.e, we used a minimal Ball-Chiu {\it Ansatz}.  As we show below, this has
the virtue of simplifying the integral equations.  We did include this term
for a single value of $\tau$ and a number of values of $b$ in Eq.~(\ref{PiA})
and found, as shown below, that it generates a small quantitative change in
some of the characteristic quantities calculated in the model but does not
alter its qualitative features.

\subsection{Wick Rotation and Euclidean Space}
The model is now almost completely specified.  Up to this point, however, we
have not considered the possible complications that may arise in employing a
Wick rotation to obtain the DSE in Euclidean space where a numerical solution
is most easily obtained.  In all model fermion DSEs that have been studied so
far the Wick rotation is not allowed in the sense that the rotation of the
$k_0$ contour always encounters at least a pole and very often a branch
cut.~\cite{Axel93,WickR} Indeed, in some models it is not possible to rotate
the contour at all.~\cite{BRW92} This point is discussed in
Refs.~\cite{CDR93,RWK92}.  We shall simply complete the definition of our
model DSE by {\bf specifying} it in Euclidean space, with metric
\mbox{$\delta_{\mu\nu}={\rm diag}(1,1,1,1)$} and with $\gamma_\mu$ hermitian:
\begin{equation}
 \Sigma(p)= \case{4}{3} g^2 \int^{\Lambda} \frac{d^4p}{(2\pi)^4} \gamma^\mu
S(k) \Gamma^\nu (k,p) D_{\mu \nu}((p-k)^2) \label{EDSE}
\end{equation}
where
\begin{equation}
S^{-1}(p) = i\gamma\cdot p + \Sigma(p) = i\gamma\cdot p A(p^2) + B(p^2)
\end{equation}
and all of the other elements in this equation are taken to be specified by
the expressions given above evaluated at Euclidean (spacelike) values of
their arguments.

Using Eqs.~(\ref{Gprop},\ref{PiA},\ref{VA},\ref{VBC},\ref{EDSE}) we obtain
the following pair of coupled integral equations for the scalar functions
that specify the model quark propagator:
\begin{eqnarray}
A(p^2) & = & 1 + \frac{16\pi}{3}\int^\Lambda \frac{d^4k}{(2\pi)^4}
        \frac{\alpha(\tau;(p-k)^2)}{(p-k)^2}
        \frac{1}{A^2(k^2)k^2+B^2(k^2)} \times \nonumber\\
 & &  \left\{ A(k^2)\frac{A(k^2)+A(p^2)}{2} \frac{1}{p^2}
	  \left[3p\cdot k - h(p,k)\right]  \right.      \nonumber \\
   &  &   \left. - A(k^2)\Delta A(k^2,p^2)
	    \left[k^2-\frac{(k\cdot p)^2}{p^2}
	       + \frac{k\cdot p}{p^2}h(p,k)\right] \right. \nonumber \\
   & &    \left. - B(k^2)\Delta B(k^2,p^2) \frac{h(p,k)}{p^2 }
\rule{0mm}{7mm}\right\}\/, \label{Aeqn} \\
B(p^2) & = & \frac{16\pi}{3} \int^\Lambda \frac{d^4k}{(2\pi)^4} 
        \frac{\alpha(\tau;(p-k)^2)}{(p-k)^2}
        \frac{1}{A^2(k^2)k^2+B^2(k^2)} \times \nonumber\\
  & &   \left\{3B(k^2)\frac{A(k^2)+A(p^2)}{2}  \right.  \nonumber \\
   & &	\left. + \left[B(k^2)\Delta A(k^2,p^2)
-A(k^2)\Delta B(k^2,p^2)\right]h(p,k)\rule{0mm}{7mm}\right\}\/, \label{Beqn}
\end{eqnarray}
where $h(p,k)=2\left[k^2p^2-(k\cdot p)^2\right]/q^2$ and 
\mbox{$\Delta F(k,p) = [F(k^2) - F(p^2)]/[k^2 - p^2]$}.

Including the additional Curtis-Pennington term in the vertex,
Eq.~(\ref{CPGamT}), these equations are modified as follows:
\begin{eqnarray}
A(p^2) & = & {\rm RHS~of~(\ref{Aeqn})} \label{AT} \\
 &+&  \frac{16\pi}{3} \int^\Lambda \frac{d^4k}{(2\pi)^4}
        \frac{\alpha(\tau;(p-k)^2)}{(p-k)^2}
        \frac{A(k^2)\Delta A(k^2,p^2)}{A^2(k^2)k^2+B^2(k^2)}
    \frac{(k^2-p^2)}{2d(k,p)}
     \frac{3(k^2-p^2)k\cdot p}{p^2}\/,
	\nonumber\\
B(p^2) & = & {\rm RHS~of~(\ref{Beqn})} \label{BT} \\
 &+& \frac{16\pi}{3}\int^\Lambda \frac{d^4k}{(2\pi)^4}
        \frac{\alpha(\tau;(p-k)^2)}{(p-k)^2}
        \frac{B(k^2)\Delta A(k^2,p^2) }{A^2(k^2)k^2+B^2(k^2)}
    \frac{(k^2-p^2)}{2d(k,p)}
    3(k^2-p^2)~.\nonumber
\end{eqnarray}

These equations were solved numerically by iteration on a logarithmic grid of
\mbox{$x= p^2/\Lambda_{\rm QCD}^2$} and
\mbox{$y=k^2/\Lambda_{\rm QCD}^2$} points.   In doing this we ensured that
our results were independent of the seed-solution and grid choice.  Our
results were independent of the UV cutoff, which was \mbox{$\Lambda^2 =
5\times 10^8\;\Lambda_{\rm QCD}^2$}, and this value was also sufficient to
ensure that the leading-log behaviour of the mass function,
Eq.~(\ref{TrueUV}) below, had become evident.

\section{Analysis of Order Parameters}

\subsection{Quark Condensate}
We are interested in determining whether the model gluon propagator specified
by Eqs.~(\ref{Gprop}) and (\ref{PiA}) can support DCSB - a crucial feature of
QCD.  The quark condensate, which is gauge-invariant, is an order parameter
for DCSB and it is easily related to the trace of the quark propagator, which
is the focus of our DSE study:
\begin{equation}
\langle \overline{q}q\rangle_\mu = -\frac{3}{4\pi^2} 
\ln\left(\frac{\mu^2}{\Lambda_{\rm QCD}^2}\right)^{d}\,
\lim_{\Lambda^2\rightarrow \infty} \left(
\ln\left(\frac{\Lambda^2}{\Lambda_{\rm QCD}^2}\right)^{-d}\,
\int_0^{\Lambda^2}
ds\, s\,\frac{B(s)}{s A(s)^2 + B(s)^2}\right)~,
\label{cndst}
\end{equation}
where \mbox{$d=12/(33-2 n_f)$} and $\mu$ is the renormalisation point for the
condensate, which is usually fixed at \mbox{$1$ GeV$^2$}.  This is the
parameter that is used to study DCSB in lattice QCD and in many of the model
studies in the continuum.  (It is clear from Eq.~(\ref{cndst}) that an
equivalent order parameter is $B(s=0)$ since if this is zero then so is the
condensate.)

As we remarked above, Eqs.~(\ref{PiA}) and (\ref{alphatau}) in the quark DSE
ensure that the leading-log behaviour of QCD is retained in the model so
that~\cite{WKR91,QCDLL,RM90}
\begin{equation}
\label{TrueUV}
\left.B(p^2)\right|_{p^2\rightarrow\infty}
\rightarrow
-\frac{4\pi^2 d}{3}
\frac{\left(\ln\left[\mu^2/\Lambda_{\rm QCD}^2\right]\right)^{-d} 
        \langle\overline{q}q\rangle_\mu}
        {p^2 \left(\ln\left[p^2/\Lambda_{\rm QCD}^2\right]\right)^{1-d}}~.
\end{equation}
This provides another means of extracting the condensate and hence a check on
its evaluation.  

In cases for which our iterative solution procedure for the DSE converged
quickly, with relative errors of less than \mbox{$1\times10^{-6}$}, the
condensate could be obtained easily.  However, for values of $b^2$\ near a
phase transition the convergence could be extremely slow.  In those cases,
the numerical solution was examined at constant intervals through the run
(say, every 50th cycle), and the condensate evaluated in each case.  Aitken
extrapolation~\cite{Aitken} was then used to find the ``infinite-cycles''
limit.  In several cases the program was subsequently run until the solutions
had converged to within
\mbox{$1\times10^{-6}$} and the extrapolated result always matched the actual
result to within a few parts in $10^{-6}$\/.

\subsection{Critical Behavior of the Condensate}

We solved Eqs.~(\ref{Aeqn}) and (\ref{Beqn}) for values of $\ln\tau$ in the
domain \mbox{$[0.0,0.7]$} and $b^2$ in \mbox{$[0.1,1.0]$} using the minimal
Ball-Chiu vertex and we plot the condensate obtained from our solutions in
Fig.~\ref{Cplt}.  This figure shows regions of unbroken and dynamically
broken chiral symmetry.

Our numerical results suggest that the condensate rises continuously from the
transition boundary and hence that the transition is second order.  As a
consequence we assumed that the order parameter,
\mbox{$\langle\overline{q}q\rangle_\mu$}, behaves as
\begin{equation}
      \langle\overline{q}q\rangle_\mu(z) \approx C
\left(1-\frac{z}{z_c}\right)^{\beta}
\end{equation}
for $z \rightarrow z_c^-$\ (for z equal to either $\ln\tau$\ or $b^2$\/) and
extracted the critical points, $z_c$, and critical exponents, $\beta$, using
ratio-of-logs methods adapted from Gaunt and Guttmann.~\cite{GaGu} We list
these quantities in Table.~\ref{betas} and, in Fig.~\ref{Ccrv}, plot the
critical curve in the $(b^2,\ln\tau)$ plane.

We also solved Eqs.~(\ref{Aeqn}) and (\ref{Beqn}) with the Curtis-Pennington
additions, Eqs.~(\ref{AT}) and (\ref{BT}), using \mbox{$\ln\tau=0.6$}.  The
critical curve (in $b^2$) in this case is shown in Fig.~\ref{BCvsCP} along
with the minimal Ball-Chiu results for the same value of $\ln\tau$.  The
effect of the Curtis-Pennington addition is to lower the critical value of
$b^2$ but, as we show below, the critical exponent is unchanged.  This curve
illustrates the point that the qualitative features of the model are not
affected by this modification of the model quark-gluon vertex.

\subsubsection{Critical Parameters for the Chiral Phase Transition}

{}From Table~\ref{betas} we find:
\begin{eqnarray}
\label{betaa}
\beta_{\rm BC} = 0.575 & \;\; ,\;\; & \sigma_{\beta} = 0.024~.
\end{eqnarray}
We note that the critical exponent obtained with \mbox{$\ln\tau=0$} is quite
different from the others.  This is a special case since for this value the
propagator does not vanish in the infrared:
\begin{equation}
\frac{g^2}{4\pi} D(q^2) = d\pi\,\frac{\Lambda_{\rm QCD}^2}{b^4}~.
\end{equation}
If we neglect this point in our analysis then we find
\begin{eqnarray}
\label{betab}
\beta_{\rm BC} = 0.572 & \;\; ,\;\; & \sigma_{\beta} = 0.020~.
\end{eqnarray}

The results in Eqs.~(\ref{betaa}) and (\ref{betab}) are in agreement with
those of Ref.~\cite{RM90} where it is argued that \mbox{$\beta = 0.589 \pm
0.031$}.  That study used $b^2=0$ and found a critical value of
\mbox{$\ln\tau = 1.69$} which complements the results reported herein, as will
be seen in Fig.~\ref{Ccrv}.

We also calculated the critical exponent using our numerical DSE solutions
obtained with the Curtis-Pennington addition to the vertex at
\mbox{$\ln\tau=0.6$}:
\begin{eqnarray}
\label{betac}
\beta_{\rm CP} = 0.579 & \;\; ,\;\; & \sigma_{\beta} = 0.015~.
\end{eqnarray}
This suggests that the vertex modification does not alter the critical
exponent of the transition; a conclusion that is also supported by the
observation that the vertex used in Ref.~\cite{RM90} was not of either of the
above forms but was, effectively, a simple modified rainbow approximation:
\begin{equation}
\Gamma_\mu(k,p) = A(k^2) \gamma_\mu~.
\end{equation}

\subsection{Confinement Test}

It is important to determine whether the model gluon propagator specified by
Eqs.~(\ref{Gprop}) and (\ref{PiA}) leads to quark confinement; i.e., the
absence of free quarks in the QCD spectrum.  This form of gluon propagator
has been constructed so that the dominant IR behaviour ensures that it
vanishes at $q^2=0$.  In order to determine whether the quark propagator
obtained as a solution to our DSE can represent a confined particle we follow
Ref.~\cite{HRM92} and adapt a method commonly used in lattice QCD to estimate
bound state masses.

We write 
\begin{equation}
\sigma_S(p^2) = \frac{B(p^2)}{p^2 A(p^2)^2 + B(p^2)^2}
\end{equation}
and define 
\begin{equation}
\Delta_S(T,\vec{x}) = \int\,\frac{d^4p}{(2\pi)^4} 
{\rm e}^{i(p_4T+\vec{p}\cdot\vec{x})} \sigma_S(p^2)~.
\end{equation}
This is the scalar part of the Schwinger function of the model quark
propagator.  If we now define
\begin{equation}
\Delta_S(T) = \int\, d^3x\,\Delta_S(T,\vec{x})
\end{equation}
and, for notational convenience,
\begin{equation}
E(T) = -  \ln\left[\Delta_S(T)\right]
\end{equation}
then it follows that if there is a stable {\it asymptotic state} with the
quantum numbers of this Schwinger function then
\begin{equation}
\label{ConfT}
\lim_{T\rightarrow\infty} 
 \frac{dE(T)}{dT} \, = m~;
\end{equation}
where $m \geq 0$ is the mass of this excitation; i.e, this limit yields the
dynamically generated quark mass.  A finite value of $m$ indicates that the
quarks are not confined since it ensures that the cluster decomposition
property is satisfied by this Schwinger function.~\cite{RWK92} 

We argue, therefore, that if the limit in Eq.~(\ref{ConfT}) exists for a
given propagator then the associated excitation {\bf is not} confined: this
is our definition of a non-confining propagator.  

In order to illustrate this point we note that the calculation of the
``constituent quark mass'' in the Nambu--Jona-Lasinio model~\cite{NJL} can be
understood in just this fashion: in this model $m$, as define above, is finite
and quarks are not confined.

In contrast, one can consider the model of Ref.~\cite{BRW92}.  Applied to
this model one finds
\begin{equation}
\frac{dE(T)}{dT} \stackrel{T\rightarrow\infty}{\sim} \kappa T
\end{equation}
where $\kappa$ is a constant and hence the limit in Eq.~(\ref{ConfT}) does
not exist.  In this case the Schwinger function does not satisfy the cluster
decomposition property and hence the quarks are confined.~\cite{RWK92}
Alternatively one may say that through self interaction the quark acquires an
infinite dynamical-mass.  This provides another way of understanding the
claim that the model of Ref.~\cite{BRW92} is confining.

Another application of this method, which is of direct interest here, is the
IR vanishing gluon propagator of Eq.~(\ref{SProp}).  In this case one has a
boson propagator and finds for the analogue of $\Delta_S(T)$:
\begin{equation}
\label{BosonD}
\Delta(T) \propto \frac{1}{ b \surd 2}
        \exp\left(-\frac{b\, T}{\surd 2}\right)
        \left(\cos\left(\frac{b\, T}{\surd 2}\right)
                - \sin\left(\frac{b\, T}{\surd 2}\right)\right)~.
\end{equation}
We remark that the Schwinger function in this case is not positive definite,
which is an easily identifiable signal in $\Delta(T)$ that is due to the pair
of complex conjugate poles, and this violates the axiom of reflection
positivity.  It follows from this that Eq.~(\ref{BosonD}) describes a field
with a complex mass spectrum and/or residues that are not positive.  This is
appropriate for particles that decay and forms the basis of the
argument~\cite{Zw91,St86,St90a,St90b} that a propagator of the type in
Eq.~(\ref{SProp}) allows coloured states to exist only for a finite time (of
the order of $1/b$) before hadronising; i.e., that the propagator describes
confined gluons.

\subsubsection{Confinement and Dressed-quark-masses}

In applying this method here it is obvious that numerical evaluation of the
Fourier transforms required in using Eq.~(\ref{ConfT}) will be hindered by
numerical noise as $T$ is increased.  In order to minimise the effect of this
noise on the derivative, we fitted $\Delta_S(T)$ to a form
\begin{equation}
\label{expfit}
C \exp\left(-m T\right)
\end{equation}
and extracted the derivative from this fit.  (Importantly, we found no
indication of the structure suggested by Eq.~(\ref{BosonD}) in our results.)
This was particularly useful with the propagators obtained using small values
of $b$ which had large dynamical masses (as one would expect since the
condensate is large in this case) and hence a rapid decline with $T$.

We applied the confinement test in the following cases: 1) The propagators
obtained with \mbox{$\ln\tau=0.1$} and $b^2$ in the range \mbox{$[0.1,1.0]$};
2) The propagator obtained with \mbox{$\ln\tau = 0$} and \mbox{$b^2 = 0.35$}
which yields the largest value of
\mbox{$-\langle\overline{q}q\rangle_\mu$} on the \mbox{$(b^2,\ln\tau)$} domain
considered; 3) Two propagators obtained with 
\mbox{$(b^2,\ln\tau) = (0.1,0.6)$} - one using the Ball-Chiu vertex and
another using the Curtis-Pennington addition.  The results obtained by
fitting the form Eq.~(\ref{expfit}) to our numerical output are presented in
Table.~\ref{TPMs}.  
In Fig.~\ref{FMF} we present plots of \mbox{$E'(T)$} for the family of
propagators obtained with \mbox{$\ln\tau = 0.1$} and this clearly illustrates
that an unambiguous determination of the dressed-quark-mass is possible in
this model.  It will be observed that the mass decreases with increasing
$b^2$.  This is easily understood in terms of the chiral phase transition: as
$b$ increases beyond $b_c$ there is no DCSB and massless current quarks
remain massless.  Since the behaviour of all the other solutions we obtained
was qualitatively the same as that described by the results presented in
Table.~\ref{TPMs} and Fig.~\ref{FMF} we infer that the model considered here
does not yield a confining quark propagator.  We also note that the rainbow
approximation studies of the fermion DSE with Eq.~(\ref{SProp}) in
Ref.~\cite{Axel93}, which address the question of confinement by a direct
continuation to Minkowski momentum space, found a quark propagator with a
pole at timelike $p^2$; i.e, a non-confining propagator.

We remark that, within numerical noise, the Curtis-Pennington addition made
no difference to the dressed-quark-mass value extracted in the cases
considered and only slightly reduced the normalisation constant $C$. Clearly,
the Curtis-Pennington addition leads only to a minor quantitative effect in
this part of our study too.

\section{Discussion and Conclusion}

It has been suggested~\cite{St90b} that the model gluon propagator in
Eq.~(\ref{SProp}) would lead to a confining quark propagator of the form
\begin{equation}
\label{FStingl}
\frac{i\gamma\cdot p + c_0}{
(i\gamma\cdot p + c_1)\,(i\gamma\cdot p + c_1^{\ast})}
\end{equation}
where $c_0\in {\Bbb R}$ and $c_1\in {\Bbb C}$ are constants; i.e., to a
fermion propagator with complex conjugate poles just as the gluon
propagator has; confinement being realised through the absence of poles
on the real timelike axis.  As Eq.~(\ref{BosonD}) shows, such a
propagator has a characteristic signature in \mbox{$\Delta_S(T)$} which
we do not see in Fig.~\ref{FMF}.  In our model then it is clear that a
fermion propagator of the type in Eq.~(\ref{FStingl}) does not arise.
This does not eliminate the possibility that it can arise in the
approach of Refs.~\cite{St86,St90a,St90b}, however, since the
rational-polynomial {\it Ans\"{a}tze} employed for the vertex functions
therein may lead to a completely different quark-gluon vertex to that
used here.  We simply remark that our results suggest that a fermion
propagator of the type in Eq.~(\ref{FStingl}) cannot arise if the
quark-gluon vertex is free of kinematic singularities.

In conclusion, we have studied a model DSE (Dyson-Schwinger equation) for the
quark propagator using a model gluon propagator that vanishes as \mbox{$q^2
\rightarrow 0$}, Eq.~(\ref{SProp}), and a light-cone-regular model
quark-gluon vertex, Eqs.~(\ref{VA}-\ref{CPdkp}).  This is the first study to
analyse the phenomenological implications of Eq.~(\ref{SProp}) in the
framework of the fermion DSE.  Our results suggest that this model can only
support DCSB for values of $\ln\tau$ and $b^2$ less than certain critical
values (see Fig.~\ref{Ccrv}) and does not confine quarks.  Qualitatively
similar results are obtained in Ref.~\cite{Axel93}.  As a consequence we
believe that this model gluon propagator is unlikely to be a useful
foundation for a chiral-dynamical model of QCD of the general coupled
Dyson-Schwinger--Bethe-Salpeter equation type considered in Refs.~\cite{CDM}.
Indeed, taken in a broader context, our results do not support the contention
that Eq.~(\ref{SProp}) is the correct form of the quark-quark interaction at
small $q^2$ in QCD.

\acknowledgements
Some of the calculations described herein were carried out using a grant of
computer time and the resources of the National Energy Research Supercomputer
Center. The work of CDR was supported by the US Department of Energy, Nuclear
Physics Division, under contract number W-31-109-ENG-38.  The work of FTH and
AGW was supported in part by the US Department of Energy under contract
number DE-FG05-86ER40273 and the Florida State University Supercomputer
Computations Research Institute which is partially funded by the US
Department of Energy through contract number DE-FC05-85ER250000. The work of
AGW was also supported by the Australian Research Council.  FTH wishes to
thank Bryan Gorman for the series convergence-acceleration code {\tt
xtrap.c}, and Drs.  Mark Novotny and Jooyoung Lee for several discussions on
phase transitions and critical parameter extraction.  CDR gratefully
acknowledges useful conversations with Axel Bender.


\begin{figure}
\caption{ Criticality plot for
  $-\langle \bar{q}q \rangle_\mu^{\frac{1}{3}}$\ as a
  function of $\ln \tau$\ and $b^2$\/.  The condensate, 
  $-\langle \bar{q}q \rangle_\mu^{\frac{1}{3}}$, is in units of MeV,
  scaled to $\mu^2 = 1 {\rm GeV}$\/, and $b^2$\ is
  in units $\Lambda_{QCD}^2$\/; the gluon regulator $\tau$\ is
  dimensionless.  }
\label{Cplt}
\end{figure}
\begin{figure}
\caption{ Critical curve for the phase transition in the
  $(\ln\tau, b^2)$\ plane.  The asterisk is the result extracted from
Ref.~\protect\cite{RM90}}
\label{Ccrv}
\end{figure}
\begin{figure}
\caption{ Comparison of the
  $-\langle \bar{q}q \rangle_\mu^{\frac{1}{3}}$\ condensate
  curves for the minimal Ball-Chiu and Curtis-Pennington Ans\"{a}tze
  for the proper vertex.  Both curves have $\ln\tau=0.6$\/.
  Diamonds, $\diamond$, connected with solid lines are the results
  from the B-C vertex; plus-signs, $+$, connected with dashed lines
  are results from the C-P vertex.  }
\label{BCvsCP}
\end{figure}
\begin{figure}
\caption{ Dressed-quark-mass curves for the family of propagators with
  the minimal Ball-Chiu vertex and $\ln\tau=0.1$\/. The masses are in units
of \mbox{$\Lambda_{\rm QCD}$}.}
\label{FMF}
\end{figure}

\begin{table}
  \caption{The critical points and exponents extracted for various
	   values of $\ln\tau$\/; the cumulative result is
           $\beta_{\rm BC}=0.575$\/, with $\sigma_\beta=0.024$\/; excluding the
	   point with $\ln\tau=0.0$\/, $\beta_{\rm BC}=0.572$\ with
	   $\sigma_\beta=0.020$\/.}
  \setdec 0.0000
  \begin{tabular}{cccc}
    $\ln\tau$&$b^2_{\rm C}$ - Critical $b^2$ value
        &$\beta$ - Critical Exponent 
        &$\sigma_\beta$ - standard deviation in $\beta$\\
    \tableline
    \dec  0.00 &\dec 0.6439 &\dec  0.609 &\dec  0.03    \\
    \dec  0.10 &\dec 0.5448 &\dec  0.579 &\dec  0.021   \\
    \dec  0.20 &\dec 0.4642 &\dec  0.570 &\dec  0.021   \\
    \dec  0.25 &\dec 0.4278 &\dec  0.579 &\dec  0.021   \\
    \dec  0.30 &\dec 0.3932 &\dec  0.573 &\dec  0.021   \\
    \dec  0.35 &\dec 0.3601 &\dec  0.570 &\dec  0.0195  \\
    \dec  0.40 &\dec 0.3289 &\dec  0.567 &\dec  0.021   \\
    \dec  0.50 &\dec 0.2706 &\dec  0.567 &\dec  0.021   \\
    \dec  0.55 &\dec 0.2437 &\dec  0.570 &\dec  0.021   \\
    \dec  0.60 &\dec 0.2180 &\dec  0.561 &\dec  0.021   \\
    \dec  0.70 &\dec 0.1710 &\dec  0.579 &\dec  0.021   \\
  \end{tabular}
  \label{betas} 
\end{table}

\begin{table}
  \caption{Asymptotic dressed-quark-mass values for the family of
	   propagators with $\ln\tau=0.1$\/,
	   the propagator which showed maximal DCSB\ 
	   (at $\ln\tau=0$\ and $b^2 = 0.35$\/),
	   and for $\ln\tau=0.6$\/, $b^2=0.1$\ with
	   both the Ball-Chiu and Curtis-Pennington vertices.
	   }
  \setdec 0.0000
  \begin{tabular}{ccccc}
    $\ln\tau$&$b^2$&$m_{\rm free}$&$C$&Comments\\
  \tableline
    \dec .1 &\dec .25    &\dec 0.410   &\dec 0.664 & B-C vertex\\
    \dec .1 &\dec .3     &\dec 0.354   &\dec 0.650 & ''\\
    \dec .1 &\dec .35    &\dec 0.296   &\dec 0.633 & ''\\
    \dec .1 &\dec .4     &\dec 0.237   &\dec 0.624 & ''\\
    \dec .1 &\dec .45    &\dec 0.176   &\dec 0.619 & ''\\
    \dec .1 &\dec .475   &\dec 0.143   &\dec 0.619 & ''\\
    \dec .1 &\dec .49    &\dec 0.122   &\dec 0.619 & ''\\
    \dec .1 &\dec .5     &\dec 0.107   &\dec 0.621 & ''\\
    \dec .1 &\dec .51    &\dec 0.0913  &\dec 0.619 & ''\\
    \dec .1 &\dec .52    &\dec 0.0739  &\dec 0.619 & ''\\
    \dec .1 &\dec .525   &\dec 0.0644  &\dec 0.619 & ''\\
    \dec .1 &\dec .53    &\dec 0.0539  &\dec 0.619 & ''\\
    \dec .1 &\dec .535   &\dec 0.0421  &\dec 0.619 & ''\\
    \dec .1 &\dec .5375  &\dec 0.0353  &\dec 0.619 & ''\\
    \dec .1 &\dec .539   &\dec 0.0308  &\dec 0.619 & ''\\
    \dec .1 &\dec .54    &\dec 0.0275  &\dec 0.619 & ''\\
  \tableline
    \dec .0 &\dec .35    &\dec 0.406   &\dec 0.667 & ''\\
  \tableline
    \dec .6 &\dec 0.1    &\dec 0.210   &\dec 0.648 & B-C vertex\\
    \dec .6 &\dec 0.1    &\dec 0.210   &\dec 0.507 & C-P vertex\\
  \end{tabular}
  \label{TPMs} 
\end{table}
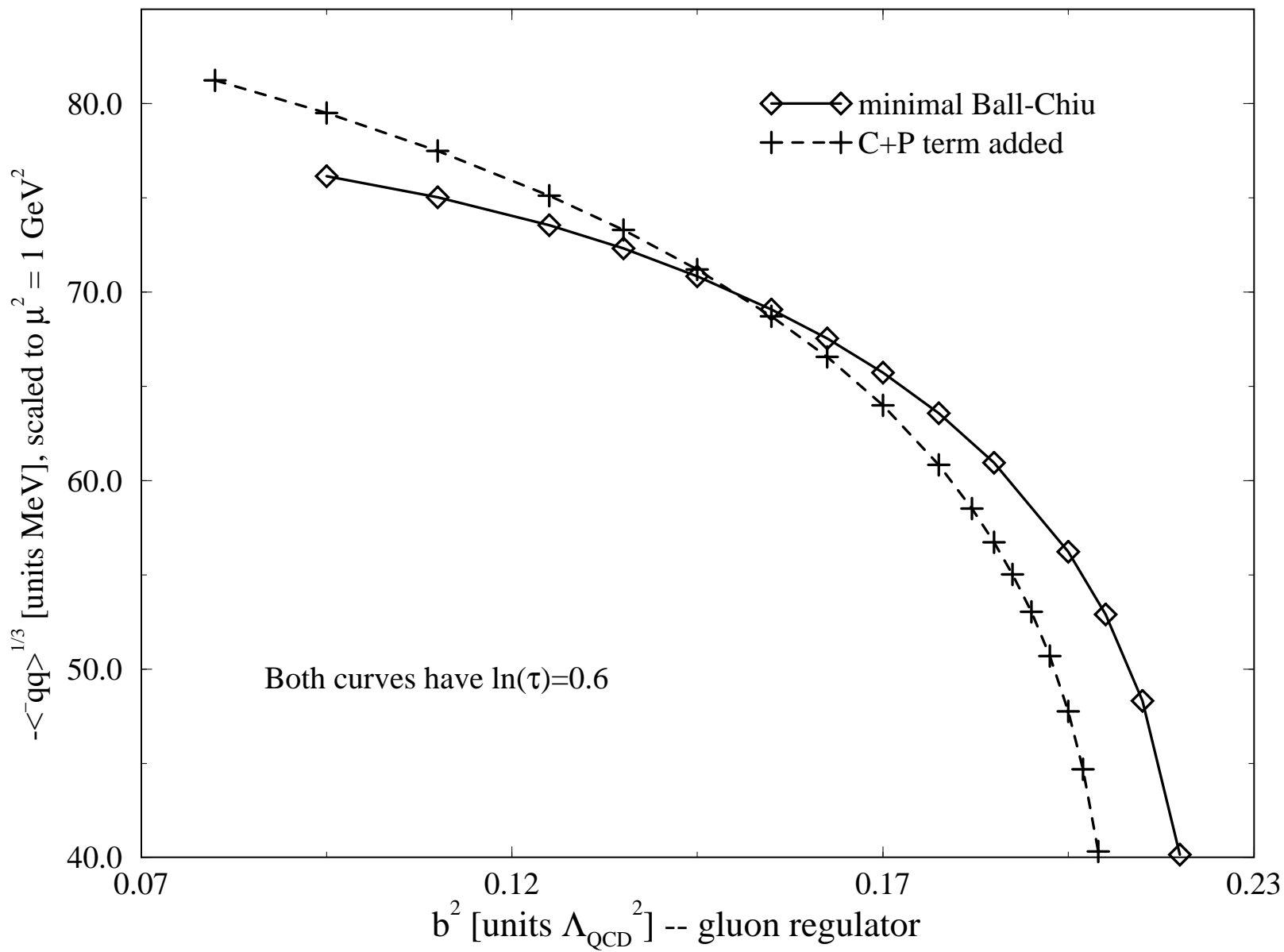
\end{document}